\documentclass[fleqn,10pt]{wlscirep}

\usepackage{graphicx}
\usepackage{amsmath}
\usepackage{multirow}
\usepackage{color}
\usepackage{bm}
\usepackage{lipsum} % for wide table
\usepackage{caption}
\usepackage{subcaption} % for subfigure
\usepackage{etoolbox}

\title{Machine learning reveals orbital interaction in crystalline materials}

\author[1, 5]{Pham Tien Lam}
\author[2, 5]{Hiori Kino}
\author[1,2]{Kiyoyuki Terakura}
\author[2, 3, 5]{Takashi Miyake}
\author[4,6]{Ichigaku Takigawa}
\author[2, 7, 8]{Koji Tsuda}
\author[1, 2, 4, *]{Dam Hieu Chi}
\affil[1]{Japan Advanced Institute of Science and Technology, 1-1 Asahidai, Nomi,Ishikawa 923-1292, Japan}
\affil[2]{Center for Materials research by Information Integration, Research and Services Division of Materials Data and Integrated System, National Institute for Materials Science 1-2-1 Sengen, Tsukuba, Ibaraki 305-0047, Japan}
\affil[3]{CD-FMat, AIST, 1-1-1 Umezono, Tsukuba 305-8568, Japan}
\affil[4]{JST, PRESTO, 4-1-8 Honcho, Kawaguchi, Saitama, 332-0012, Japan}
\affil[5]{ESICMM, National Institute for Materials Science 1-2-1 Sengen, Tsukuba, Ibaraki 305-0047, Japan}
\affil[6]{Graduate School of Information Science and Technology, Hokkaido University, N-14, W-9, Sapporo 060-0814, Japan}
\affil[7]{Department of Computational Biology and Medical Sciences, Graduate School of Frontier Sciences, University of Tokyo, 5-1-5 Kashiwanoha, Kashiwa, Japan}
\affil[8]{RIKEN Center for Advanced Intelligence Project, 1-4-1 Nihombashi Chuo-ku, 103-0027 Tokyo, Japan}
\affil[*]{dam@jaist.ac.jp}

\begin{abstract}
We propose a novel representation of crystalline materials named orbital-field matrix (OFM) based on the distribution of valence shell electrons. We demonstrate that this new representation can be highly useful in mining material data. Our experiment shows that the formation energies of crystalline materials, the atomization energies of molecular materials, and the local magnetic moments of the constituent atoms in transition metal--rare-earth metal bimetal alloys can be predicted with high accuracy using the OFM. Knowledge regarding the role of coordination numbers of transition metal and rare-earth metal elements in determining the local magnetic moment of transition metal sites can be acquired directly from decision tree regression analyses using the OFM.  
\end{abstract}
\begin{document}

\flushbottom
\maketitle
\thispagestyle{empty}

%\noindent Please note: Abbreviations should be introduced at the first mention in the main text – no abbreviations lists. Suggested structure of main text (not enforced) is provided below.

\section*{Introduction}

Recently, an increasing volume of available experimental and quantum-computational material data along with the development of machine-learning techniques has opened up a new opportunity to develop methods for accelerating discoveries of new materials and physical chemistry phenomena. By using machine-learning  algorithms, hidden information of materials, including patterns, features, chemical laws, and physical rules, can be automatically discovered from both first-principles-calculated data and experimental data \cite{data_mining_materials_science_PRB2012, identifying_zeolite_framework_JPC_C2012, find_missing_ternary_oxide_chem_mater_2012, find_DFT_PRL_2012, Materials_cartography_chem_mater_2015, PhysRevLett_big_data_materials_descriptors_sheffer, JCP_parallel_lasso_SMM, JCP_LMM}. It is common knowledge that, in a material dataset, the most important information for identifying a material is its structure. Information on the structure of a material is usually described using a set of atoms with their coordinates and periodic unit-cell vectors, which are required for crystalline systems. From the viewpoint of data science, the material data using this primitive representation can be categorized as unstructured data, and the mathematical basis on such material data is only the algebra of sets. Therefore, advanced quantitative machine-learning algorithms can hardly be applied directly to conventional material data owing to the limitation of the algebra of the primitive data representation.

In order to apply well-established machine-learning methods including predictive learning and descriptive learning, it is necessary to convert the primitive representation into vectors or matrices such that the comparison and calculations using the new representation reflect the nature of materials and the actuating mechanisms of chemical and physical phenomena. Various methods for encoding materials have been developed in the field of materials informatics. Behler and coworkers \cite{BehlerPRL, BehlerJCP, Nongnuch_Artrith_nanoparticles, Eshet, Eshet1, Artrith, Artrith1} utilized atom-distribution-based symmetry functions to represent the local chemical environment of atoms and employed a multilayer perceptron to map this representation to the associated atomic energy. The arrangement of structural fragments has also been used to encode materials to predict the physical properties of molecular and crystalline systems \cite{Pilania, Materials_cartography_chem_mater_2015}. Isayev used band structure and density of states (DOS) fingerprint vectors as a representation of materials to visualize material space \cite{Materials_cartography_chem_mater_2015}. Rupps and coworkers developed a descriptor known as the Coulomb matrix (CM) for the prediction of atomization energies and formation energies \cite{Rupps, Faber_Coulomb_matrix, Rupp_tutorial}. Although the CM is very successful in predicting of properties of molecules, its performance with regard to the the formation energies of crystal systems is relatively poor\cite{Faber_Coulomb_matrix}. These representations do not include explicit information about the atomic orbitals or the nature of chemical bonding in materials, which is necessary for the determination of the electronic structure and the resulting physical properties, so the learning results have low interpretability in the language of physical chemistry. To study materials using machine-learning approaches, both the accuracy and the interpretability of the learnt models are important aspects. \cite{Merckt95aboutbreaking_accuracy_vs_interpretability}. To render data-driven approaches meaningful and useful for materials science studies, it is necessary to design material representations with which the results derived using machine-learning methods can be interpreted in the language of physical chemistry. Further, structural information on the materials should be included explicitly in the learning results for supporting the materials design processes. In this paper, with emphasis on the interpretability of the derived learning results, we propose a novel representation of materials by utilizing domain knowledge in encoding them.

It has been well established in fundamental chemistry that certain important aspects of the electronic structure can be deduced from a simple description of the nearest valence electrons around an atom in a molecule or crystal system, e.g., the Lewis theory provides powerful tools for studying the structure of molecules \cite{chemistry_chemcial_reactivity}. The ligand field theory is another example of a theory developed based on this intuition, and several fruitful results have been obtained using this theory \cite{molecular_orbital_of_transition_metal_complexes}. In this work, we utilize this domain knowledge in encoding the materials to propose a novel representation of materials named orbital-field matrix (OFM) by using the coordination of valence orbitals (electrons). A material or a local structure is encoded by counting the valence orbitals of the nearest neighbors. We focus on magnetic materials based on rare earth--transition metal (RT) alloys and RT alloys including a light element X, which may be B, C, N, or O (RTX). 
 
For verifying the applicability of the proposed material representation, we first examine the decision tree for predicting the magnetic moment of Mn, Fe, Co, and Ni in RT alloys. The decision trees learnt from the RT alloy data show that the coordination numbers of the occupied $d$ orbitals of transition metals and occupied $f$ orbitals of rare-earth metals play an important role in determining the local magnetic moment of the transition metal sites. The obtained results confirm the interpretability of our OFM representation in terms of structural, physical chemistry.  Kernel ridge regression (KRR) analyses using standard techniques and similarity measures are carried out in learning prediction models to predict the local magnetic moments and formation energies of the alloy materials. Our computational experiments show that the OFM representation can accurately reproduce the DFT-calculated local magnetic moments of transition-metal sites in RT alloys, formation energies of crystalline systems, and atomization energies of molecular systems. The high prediction accuracy confirms the practicability of our OFM representation.

\section*{Methodology}
\subsection*{Representation of materials}
\begin{figure}[]
    \centering
    \includegraphics[width=0.8\textwidth]{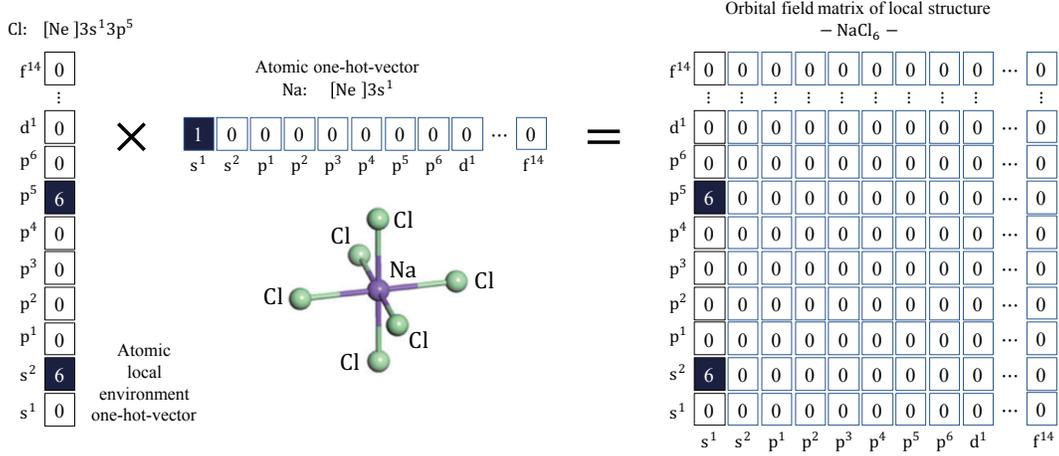}
    \caption{\label{fig:NaCl-representation-1} OFM representation for a Na atom in an octahedral site surrounded by 6 Cl atoms: atomic one-hot-vector for Na (middle), representation for the 6 Cl atoms surrounding the Na atom (left), and representation for the Na atom surrounded by 6 Cl atoms (right). }
\end{figure}

For designing the representation for a material, we start with the representation for an atom as a building block of the material. We utilize the standard notation for electron configuration to develop the representation for an atom, e.g., the electron configurations of Na and Cl are [Ne]$3s^1$ and [Ne]$3s^23p^5$, respectively. In order to convert this standard notation into a numerical vector, we borrow the idea of one-hot-vector in the field of natural language processing, in which a word is represented by a bit vector having the dimension of the number of words in a dictionary. The vector consists of elements with the values of 0, with the exception of a single element used uniquely to identify the word. The representation of an atom is then converted from the standard notation into a one-hot-vector $\vec{O}_{atom}$ by using a dictionary of the valence subshell orbitals: $D =  \{s^1, s^2, p^1, p^2, ..., p^6, d^1, d^2, ..., d^{10}, f^1, f^2, ..., f^{14}\}$ (e.g., $d^5$ indicates the electron configuration in which the valence $d$ orbital holds 5 electrons), which consists of 32 elements (Fig \ref{fig:NaCl-representation-1}). 

Next, we design the representation of a local chemical environment by considering the sum of the weighted vector representations of all atoms in the environment:
\begin{equation}
\vec{O}_{env} = \sum_k \vec{O}_k w_k,
\end{equation}
where $\vec{O}_k$ is the representation vector of atom $k$ and $w_k$ is the weight of this atom, which measures the contribution of the atom. An atom at site $p$ in a chemical environment can be represented using the OFM as follows: 
\begin{align}
\label{eq:eq2}
X^{(p)} = &\sum_{k \in n_p} \vec{O}_k^T  \times \vec{O}^{(p)}_{center} \times w_k, \notag\\ 
X_{ij}^{(p)} = &\sum_{k \in n_p }o^k_j o^{(p)}_i \frac{\theta_k^{(p)}}{\theta_{max}^{(p)}}
\end{align} 
where $i, j \in \{s^1, s^2, p^1, ..., p^6, d^1, ..., d^{10}, f^1, ..., f^{14}\}$, which is the set of electron configurations in valence orbitals; $k$ is the index of the nearest-neighbor atoms; $n_p$ is the number of nearest-neighbor atoms surrounding site $p$; $w_k$ is a weight that represents the contribution of atom $k$ to the coordination number of the center atom, $p$; $o^k_j$ and $o^p_i$ are elements of the one-hot-vectors of the $k^{th}$ neighboring atom and the center atom $p$ ($o^u_v$ equals 1 if the valence orbitals of the atom at site $u$ have electron configuration of type $v$, else it equals to 0) representing the electron configuration. The weight, $w_k = \theta_k^{(p)}/\theta_{max}^{(p)}$, is determined from a scheme employing the Voronoi polyhedron proposed by O'Keeffe \cite{Okeeffe_coordination_number} implemented in pymatgen code \cite{pymatgen}. In this expression, $\theta_k^p$ is the solid angle determined by the face of the Voronoi polyhedral separating the atom $k$ and the atom $p$, and $\theta_{max}^p$ is the maximum among all the solid angles determined by the face of the Voronoi polyhedral separating the atom $p$ and the nearest-neighbor atoms.

Additionally, in order to incorporate the information on the size of the valence orbitals, the distance $r_{pk}$ between the center atom $p$ and the neighboring atom $k$ should be included in the weight, $w_k$. We propose the following form for the calculation of the OFM elements:
\begin{equation}
	\label{eq:3}
  X_{ij}^{(p)} = \sum_{k \in n_p }o^k_j o^{(p)}_i \frac{\theta_k^{(p)}}{\theta_{max}^{(p)}} w(r_{pk}),
\end{equation}
where $w(r_{pk})$ is a function representing the contribution of the distance to the weight. In this work, we use the inverse of the distance as the distance-dependent weight function: $w(r_{pk}) = 1 / r_{pk}$.  

Composing the descriptor for a structure (a molecule or a crystal system) from its local structure representation requires careful consideration so that as much information as possible is included. In this work, for the atomization energy,  we simply take the sum of the descriptors of the local structures as the descriptor for the entire structure. For the average formation energy (per atom), the descriptor for the entire structure is composed averaging the descriptors of the local structures.

\begin{figure*}[]
\centering
    \begin{subfigure}[b]{0.40\textwidth}
        \includegraphics[width=1.\textwidth]{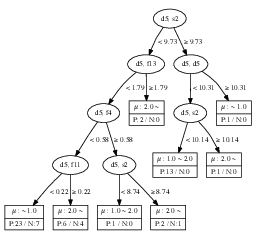}
        \caption{}
        \label{fig:Mn}
    \end{subfigure}
    \begin{subfigure}[b]{0.38\textwidth}
        \includegraphics[width=1.\textwidth]{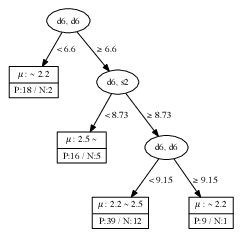}
        \caption{}
        \label{fig:Fe}
    \end{subfigure}\\
    \begin{subfigure}[b]{0.38\textwidth}
        \includegraphics[width=1.\textwidth]{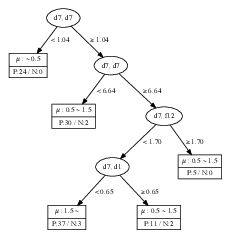}
        \caption{}
        \label{fig:Co}
    \end{subfigure}
    \begin{subfigure}[b]{0.38\textwidth}
        \includegraphics[width=1.\textwidth]{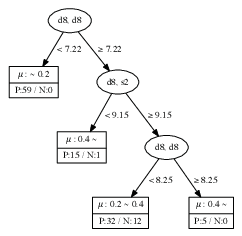}
        \caption{}
        \label{fig:Ni}
    \end{subfigure}
\caption{\label{fig:Decision_tree} Decision tree regression for Mn (a), Fe (b), Co (c), and Ni (d). In each leaf, the upper part indicates the values of local magnetic moments, whereas the lower part indicates the number of positive (P) and negative (N) examples.}
\end{figure*}

\section*{Results and discussion}

\subsection*{Prediction of local atomic properties}
We now examine how the OFM can be employed to predict the local atomic properties of materials. In this work, we focus on the local magnetic moment of transition metals in RT alloys, whose dataset includes 658 structures collected from the Materials Project database \cite{MaterialsProject, materialsAPI}. We select the structures by combining transition metals and rare-earth metals from $\{$Sc, Ti, V, Cr, Mn, Fe, Co, Ni, Cu, Zn, Y, Zr, Nb, Mo, Tc, Ru, Rh, Pd, Ag, Cd, Rh, Pd, Ag, Cd, Hf, Ta, W, Re, Os, Ir, Pt, Au$\}$ and $\{$La, Ce, Pr, Nd, Pm, Sm, Eu, Gd, Tb, Dy, Ho, Er, Tm, Yb, Lu$\}$. 

Since the local magnetic moment of a transition metal site is determined by the number of unpaired electrons of $d$-orbitals, our description of the local structure in terms of the coordination of valence electrons is expected to include a significant amount of information for predicting the local magnetic moment. We first examine which elements in the OFM determine the local magnetic moment of the Mn, Fe, Co, and Ni sites in the RT dataset through decision tree regression analyses.

As explained above, the representation of atoms and surrounding environments is derived by converting the standard notation into an one-hot-vector, using a dictionary of the valence subshell orbitals (Fig \ref{fig:NaCl-representation-1}, Equation 2). The $(d^5,d^5)$, $(d^6,d^6)$, $(d^7,d^7)$, and $(d^8,d^8)$ elements in the OFM correspond, respectively, to the coordination numbers of Mn atoms surrounding a Mn site, Fe atoms surrounding an Fe site, Co atoms surrounding a Co site, and Ni atoms surrounding a Ni site. All the atoms in an RT alloy have valence subshell $s$ orbitals occupied by 2 electrons; therefore, the $(d^5,s^2)$, $(d^6,s^2)$, $(d^7,s^2)$, and $(d^8,s^2)$ elements correspond to the total coordination number of a Mn site, an Fe site, a Co site, and a Ni site, respectively. The  $(d^5,f^n)$, $(d^6,f^n)$, $(d^7,f^n)$, and $(d^8,f^n)$ elements correspond to the coordination numbers of rare-earth atoms surrounding a Mn site, an Fe site, a Co site, and a Ni site, respectively. Moreover, $(d^n ,d^1)$ corresponds to the coordination number of either La, Ce, Ga, or Lu (among which the valence subshell $d$ orbitals are occupied by 1 electron) surrounding a transition metal site. 

The decision tree regressions for the local magnetic moment for Mn, Fe, Co, and Ni sites derived from the data are summarized in Figure 2. From these results, it is clear that $(d^n,d^n)$ elements, namely the coordination number of a transition metal site surrounded by transition metal atoms of the same kind, are important for the local magnetic moment of the Fe, Co, and Ni sites. It is interesting to note that, to obtain a high value of the magnetic moment, the $(d^n, d^n)$ elements of the OFM should be in a specific range. For instance, Fe sites tend to have a magnetic moment less than $2.2 \mu_B$ when the $(d^6,d^6)$ element is less than 6.6 or greater than 9.15. This implies that the Fe atom appears to have a smaller magnetic moment when surrounded by less than 7 Fe atoms or more than 9 Fe atoms. Further, the magnetic moment of Fe sites may be greater than $2.5 \mu_B$ when the $(d^6,d^6)$ element is greater than 6.6, but the $(d^6,s^2)$ element, namely the total coordination number including the contribution of rare-earth metal atoms, is less than 8.73. In contrast, Ni sites tend to have a small magnetic moment (less than $0.2 \mu_B$) when the $(d^8,d^8)$ element is less than 7.22, but a large magnetic moment (greater than $0.4 \mu_B$) can be obtained when the $(d^8,d^8)$ element is greater than 8.25. This implies that the Ni atom appears to have a large magnetic moment when surrounded by more than 9 Ni atoms. The magnetic moment of Ni sites may be greater than $0.4 \mu_B$ when the $(d^8,d^6)$ element is greater than 7.22, but the $(d^8,s^2)$ element, namely the total coordination number including the contribution of rare-earth metal atoms, is less than 9.15. 

For Co, the existence of rare-earth elements plays a significant role in determining the local magnetic moment, since the $(d^7,f^{12})$ and $(d^7,d^1)$ elements appear as nodes in the decision tree. This result means that a proper small amount of a rare-earth metal in which the valence subshell $d$ orbitals are occupied by 1 electron (La, Ce, Gd, Lu) may effectively increase the local magnetic moment of Co sites. The tree for Mn sites appears to be more complicated than for the other transition-metals, which can be attributed to the complicated magnetic properties of the $d^5$ configuration of Mn. The obtained decision trees obviously suggest that the coordination numbers of the occupied $d$ orbitals of transition metals and occupied $f$ orbitals of rare-earth metals play an important role in determining the local magnetic moment of the transition metal sites. This result agrees with the fact that in RT compounds, there are three types of interactions including the magnetic interaction between transition-metal (T) atoms in the T sublattices (T--T interaction), the magnetic interaction between rare-earth (R) atoms and the T sublattices (R--T interaction), and the magnetic interaction between R atoms in the R sublattices (R--R interaction). The T--T interaction dominates in RT compounds because the delocalization and spatial extent of the $3d$ electron wave functions of T atoms are much more pronounced than those of the $4f$ electrons. The R--T interaction is weak in comparison to the T--T interaction; however, the R--T interaction plays an important role in determining the magnetic structure of RT compounds. This confirms the interpretability in terms of structural, physical chemistry of the learning results from the data represented by the OMF descriptors. 

In the next step, we examine how the local magnetic moment can be represented by the OFM descriptors based on the fact that materials with higher similarity (as estimated by the descriptors) should possess similar local magnetic moments. For this purpose, we employ a simple nearest-neighbor regression method to predict the local magnetic moments, and the cross-validated RMSE is used to measure the performance of our descriptors. In the nearest-neighbor regression, a property of a data point is deduced from the properties of the nearest-neighbor points in the training data. In this work, we employ a nearest-neighbor regressor implemented in the scikit-learn package \cite{scikit-learn}. The number of nearest neighbors is fixed as 5, and the nearest neighbors are determined by a brute-force search. The prediction is weighted by the distance to the nearest neighbors.

\begin{table}[t]
\caption{Cross-validation RMSE ($\mu_B$) and the coefficient of determination $R^2$ in the prediction of the local magnetic moments obtained by nearest-neighbor regression with selected distance measurements.}
\centering
\label{tab:prediction_local_magnetic_moment}
\begin{tabular}{p{1.5cm}p{1cm}p{1cm}p{1cm}p{1cm}p{1cm}p{1cm}}
\hline
\hline
Distance        & $d_{eucl}$ & $d_{man}$ & $d_{cos}$ & $d_{bar}$ & $d_{can}$ & $d_{cor}$  \\
\hline
RMSE  & 0.26      & 0.21      & 0.23   & 0.21        & 0.21     & 0.23          \\
\hline
$R^2$           & 0.86      & 0.90      & 0.89   & 0.90        & 0.90     & 0.90          \\
\hline
\hline
\end{tabular}
\end{table}

Table \ref{tab:prediction_local_magnetic_moment} summarizes the cross-validation RMSE and the coefficient of determination $R^2$ between the observed and predicted values, obtained with our nearest-neighbor regression and different distance measurements. The results are obtained as the OFM weighted by distance (Eq. \ref{eq:3}). It may be noted that, for the prediction of the local magnetic moment, the difference between the distance-weighted and non-distance-weighted (Eq. \ref{eq:eq2}) OFM is negligible. We achieve a reasonable performance in the prediction of the local magnetic moments with an RMSE of approximately $0.2\ \mu_B$ and $R^2$ of 0.9. This result indicates that close materials in our description space of local structure yield similar local magnetic moments, which implies that our data representation includes significant information about the local magnetic moments. 

To further improve the prediction of the local magnetic moment, we apply KRR as the model for predicting the local magnetic moment. We obtain a cross-validated RMSE of $0.18 ~\mu_B$, a cross-validated MAE of $0.05 ~\mu_B$, and an $R^2$ value of $0.93$ as indicated in Table \ref{tab:krr_local_m}. 

\begin{table}
\caption{Cross-validation RMSE ($\mu_B$), cross-validation MAE ($\mu_B$), and coefficient of determination $R^2$ in the prediction of the local magnetic moments obtained by KRR regression with orbital-field-matrix (OFM) and Coulomb matrix (CM) descriptors.}

\centering
\label{tab:krr_local_m}
\begin{tabular}{p{3cm}p{2.5cm}p{2.5cm}}
\hline
\hline
Descriptor & OFM  & CM \\
\hline
RMSE       & 0.18 & 0.21 \\
MAE        & 0.05 & 0.11 \\
R$^2$      & 0.93 & 0.90 \\
\hline 
\hline
\end{tabular}
\end{table}

For comparison, we adopt the CM descriptor proposed by Rupp and coworkers \cite{Rupps} to represent the local structure of a center atom and its neighbors determined by the Voronoi polyhedra scheme. We treat the local structures in the same way as isolated molecules, and the calculated CM descriptors are used for predicting the local magnetic moments by using KRR regression. 
By using this descriptor, we obtain the a cross-validated RMSE of approximately 0.21 $\mu_B$, a cross-validated MAE of $0.11 ~\mu_B$, and an $R^2$ value of 0.90, as indicated in Table \ref{tab:krr_local_m}. 
The obtained results clearly confirm that the OFM descriptor, which includes information on the coordination of valence electrons, is more informative and, consequently, yields a better prediction accuracy than the CM descriptor for the local magnetic moment of RT alloys.

\subsection*{Prediction of material properties}

For predicting the properties of materials, it is required to develop descriptors for those materials. In this study, the descriptor for a material is built from the descriptors of its local structures. The prediction accuracy for a physical property will depend strongly on how well the descriptors for the material are composed from the descriptors of their local structures. With the aim of obtaining a prediction model with high prediction accuracy, the representation of materials is usually designed to include as much information as possible, with a large number of descriptors, without considering their interpretability. In this work, we focus more on developing descriptors taking into consideration of both the applicability and interpretability. Therefore, instead of designing a complicated representation for materials, we choose a simple approach in which the descriptor of a material is derived by simply averaging or summing the descriptors for the local structures of its constituent atoms. We implement the prediction models for the formation energies of crystalline systems and the atomization energies of molecular systems to examine the applicability of OFM descriptors.

\begin{figure}[]
    \centering
    \includegraphics[width=0.52\textwidth]{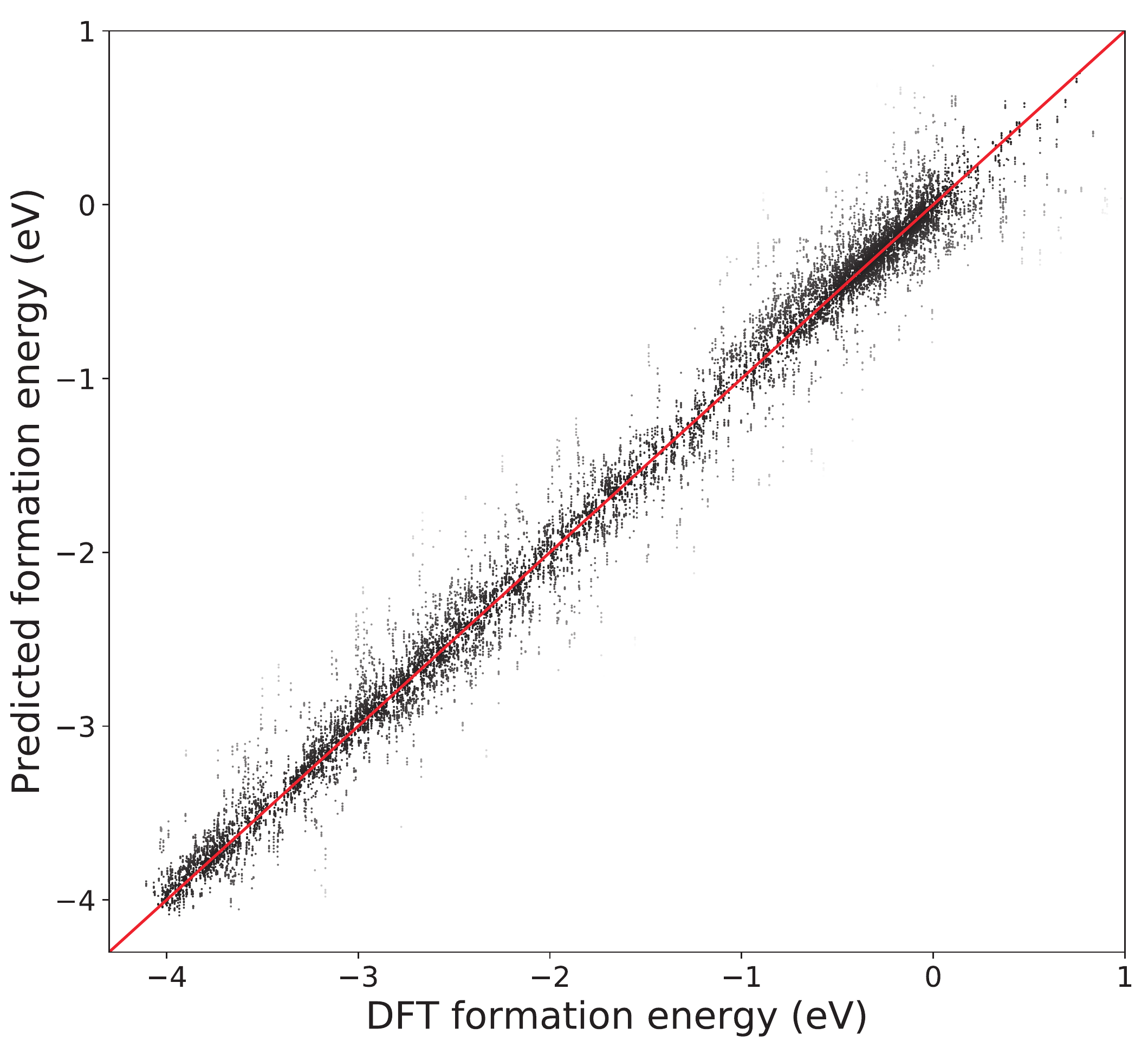}
    \caption{\label{fig:FormationEnergies} Comparison of the formation energies calculated using DFT and those predicted through machine-learning, using OMF.}
\end{figure}

For crystalline systems, we focus on transition-metal binary alloys, TT, and rare-earth--transition-metal alloys, RT, as well as RTX and TTX, which are RT and TT alloys that include a light element X = B (RTB), C (RTC), N (RTN), or O (RTO). We select the transition metals from $\{$Sc, Ti, V, Cr, Mn, Fe, Co, Ni, Cu, Zn, Y, Zr, Nb, Mo, Tc, Ru, Rh, Pd, Ag, Cd, Rh, Pd, Ag, Cd, Hf, Ta, W, Re, Os, Ir, Pt, Au$\}$, the rare-earth metals from $\{$La, Ce, Pr, Nd, Pm, Sm, Eu, Gd, Tb, Dy, Ho, Er, Tm, Yb, Lu$\}$, and X from $\{$B, C, N, O$\}$. We collect the data of more than four thousand compounds including their structures and formation energies from the Materials Project repository: 1510 RTX compounds, 1311 TTX compounds, 692 RT compounds, and 707 TT compounds. We use the average of the descriptors for their local structures to build the descriptor for these materials.

For comparison, we also implement the CM descriptors for these crystalline systems based on the Ewald sum developed by Faber and coworkers \cite{Faber_Coulomb_matrix}. We use a KRR model with a Laplacian kernel for both OFM and CM descriptors. The 10-fold cross-validated comparison between the DFT calculated formation energies and the ML predicted formation energies is shown in Fig. \ref{fig:FormationEnergies}. The DFT-calculated and ML-predicted formation energies show good agreement with an $R^2$ value of 0.98, a cross-validated RMSE of 0.19 eV/atom, and cross-validated MAE of 0.11 eV/atom. This result is better than that obtained using CM with an $R^2$ value of 0.87, a cross-validated RMSE of 0.47 eV/atom, and a cross-validated MAE of 0.39 eV/atom, as summarized in Table. \ref{tab:FormationEnergies}. 

\begin{table}[]
\centering
\caption{Cross-validation RMSE (eV/atom), cross-validation MAE (eV/atom), and coefficient of determination $R^2$ for RTX and the QM7 dataset by using orbital-field matrix (OFM) and Coulomb matrix (CM) descriptors}
\label{tab:FormationEnergies}
\begin{tabular}{p{2cm}p{1.5cm}p{1.5cm}p{1.5cm}p{1.5cm}}
\hline\hline
Dataset        & \multicolumn{2}{c}{RTX}    & \multicolumn{2}{c}{QM7} \\
\hline
Descriptor     & OFM       & CM \cite{Faber_Coulomb_matrix} & OFM    & CM \cite{Rupps} \\
\hline
RMSE  & 0.190      & 0.470           & 0.043  & 0.040           \\
MAE            & 0.112 & 0.390      & 0.027  & 0.020           \\
$R^2$          & 0.98      & 0.87           & 0.98   & 0.99           \\
\hline\hline
\end{tabular}
\end{table}

For molecular systems, we focus on atomization energies of organic molecules. We use the QM7 dataset with 7195 organic molecules \cite{Rupps, blum_qm7}. The descriptor of a molecule is built by summing over the descriptors of its local structures. By using our OFM representation and KRR regression, we obtain a cross-validated RMSE of 0.043 eV/atom, a cross-validated MAE of 0.027 eV/atom, and an $R^2$ value of 0.98, whereas the CM yields a cross-validated RMSE of 0.040 eV/atom, a cross-validated MAE of 0.020 eV/atom, and an $R^2$ value of 0.99 \cite{Rupps, Faber_Coulomb_matrix, Rupp_tutorial}, as indicated in Table. \ref{tab:FormationEnergies}. 

\begin{figure*}[]
\centering
    \begin{subfigure}[b]{0.34\textwidth}
        \includegraphics[width=1.\textwidth]{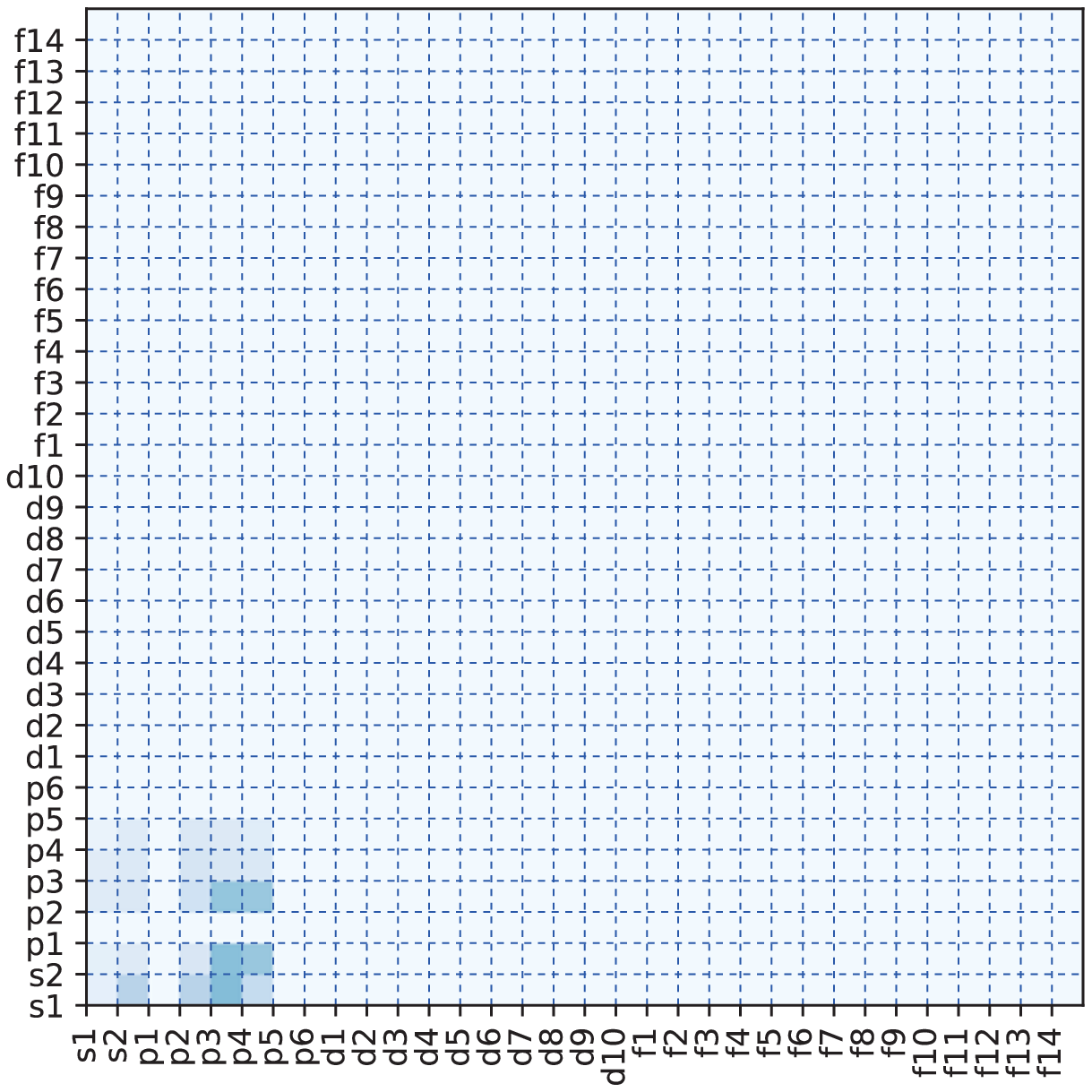}
        \caption{}
        \label{fig:qm7_std}
    \end{subfigure}
    \begin{subfigure}[b]{0.4\textwidth}
        \includegraphics[width=1.\textwidth]{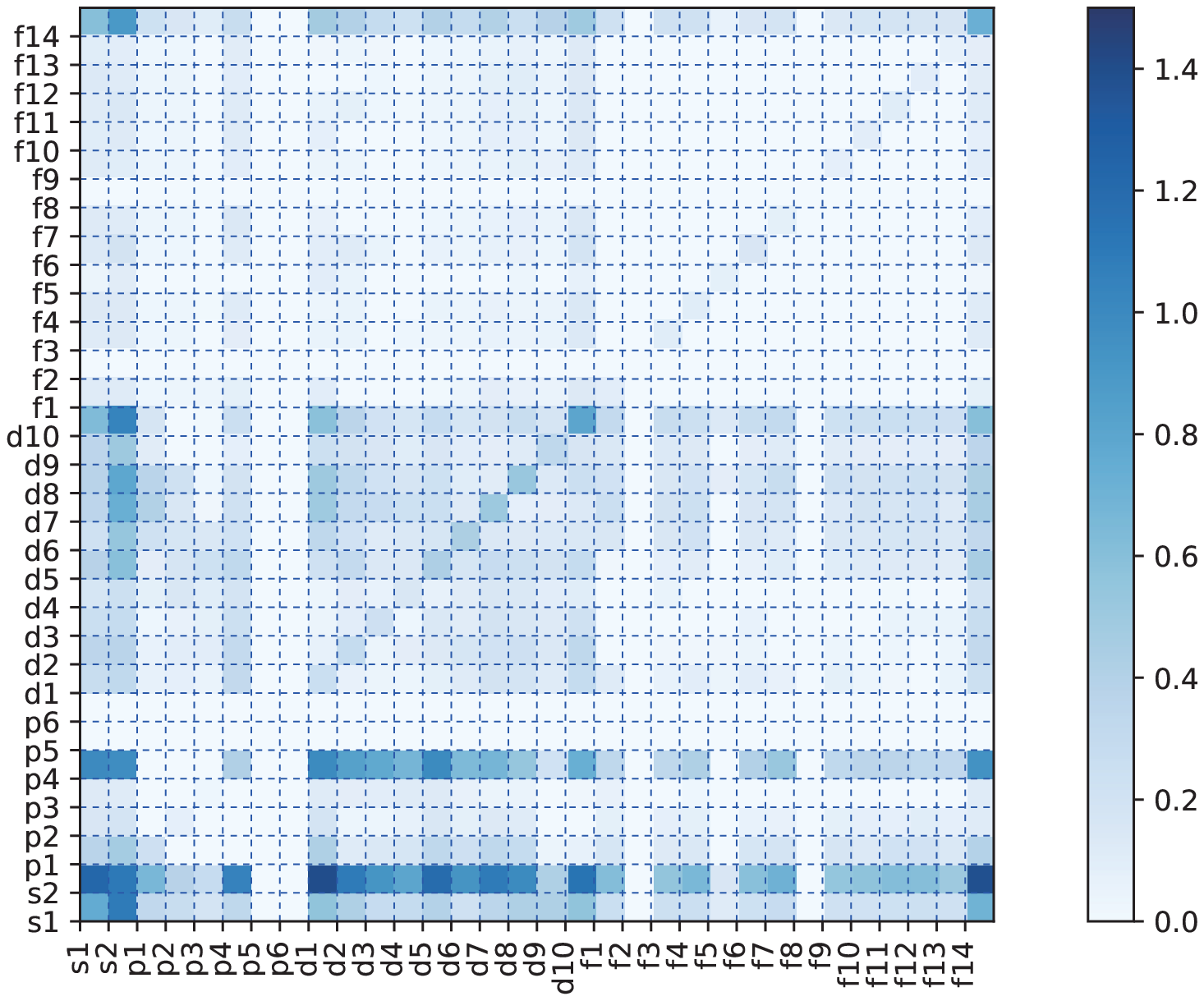}
        \caption{}
        \label{fig:rtx_std}
    \end{subfigure}
    
\caption{\label{fig:qm7_rtx_std} Standard deviation of local OFM of QM7 (a) and RTX (b) datasets.}
\end{figure*}

This result confirms that the construction of the OFM of a material, by averaging or summing the descriptors of all the local structures of the constituent atoms, yields a better prediction accuracy than the CM descriptor for the formation energy of RTX systems, and a comparable to the CM descriptor for the atomization energy of organic molecular systems in the QM7 dataset. It may be noted that for molecular systems (i.e., the QM7 dataset only contains the light elements such as C, H, O, N, and S), the CM descriptor yields a slightly better result than our OFM.  However, for RTX systems with a variety of elements (i.e., the RTX dataset contains transition-metals, rare-earth metals, and light elements),  our OFM shows a superior prediction ability. To capture the difference in the complexity of the QM7 and RTX datasets, we calculate the standard deviation of the OFM of all local structures for each dataset.  Fig. \ref{fig:qm7_rtx_std}, shows a comparison between the QM7 and RTX datasets. It is clearly seen that the QM7 dataset contained the only a small number of non-zero OFM elements at the lower left of Fig. \ref{fig:qm7_rtx_std} (a), whereas the RTX dataset exhibits a large variety of OFMs, \ref{fig:qm7_rtx_std} (b). Moreover, the QM7 dataset presents a small deviation of the OFM, and the RTX dataset has a larger deviation. This implies that the RTX dataset has higher diversity in both composition and structure than the QM7 dataset. This result indicates that the OFM may be used not only for learning properties of both crystalline and molecular systems with large diversity in atomic composition and structure, but also for studying structural properties of materials. 

\section*{Conclusion}

We have proposed a novel representation of crystalline materials named orbital-field matrix (OFM) based on the distribution of valence shell electrons. We demonstrated that this new representation can be highly useful in describing and measuring the similarities of materials or local structures in transition metal--rare-earth metal bimetal alloys. Our experiments show that our OFM can accurately reproduce the DFT-calculated local magnetic moment of transition sites in RT alloys with a cross-validated RMSE of 0.18 $\mu_B$ and an $R^2$ value of 0.93. Moreover, it can be interpreted in the language of physical chemistry, that is the ligand field theory in the local magnetic moment. The decision tree regression shows the importance of the coordination numbers of the occupied $d$ orbitals of transition metals and occupied $f$ orbitals of rare-earth metals in determining the local magnetic moment of the transition metal sites. The formation energies of crystalline systems and atomization energies of molecular systems can be well predicted using our OFM. With KRR representation, the formation energies of the crystalline systems and atomization energies of molecular systems can be accurately reproduced with an $R^2$ of approximately $0.98$. With the information about the coordination of the atomic orbitals, the OFM shows a superior applicability for the systems with high diversity in atomic composition and structure. The acquired results suggest that OFM could be useful in mining chemical/physical information of materials from available datasets by using modern machine-learning algorithms. 

\section*{Computational details}

\subsection*{First-principles calculation}
We employed VASP 5.4.1 \cite{vasp1,vasp2,vasp3,vasp4} with the GGA/PBE exchange-correlation functional\cite{ggapbe1, ggapbe2} to calculate the local magnetic moments of these structures. We followed the Materials Project database on the choice of the PAW projectors\cite{paw1,paw2}, and employed pymatgen 4.3.0 \cite{pymatgen} to prepare the VASP input files with the gaussian smearing o f 0.1~eV of MITRelaxSet and the k-point mesh density of 150~$\AA^{-3}$. The systematic simulations in this study were assisted by OACIS\cite{oacis}.

\subsection*{Machine learning}
Parameters $\gamma$ and $\lambda$ are determined in an inner loop of the 10-fold cross validation by using a logarithmic-scale grid to predict the local magnetic moment.
We optimize the hyperparameters of the KRR model to predict formation energies, kernel width $\sigma$ and regularization parameter $\lambda$, by minimizing the 10-fold cross-validated RMSE. The optimized parameters are identified by searching over 2500 pairs of $\sigma$ and $\lambda$ on a 2D logarithmic grid.
These procedures are routinely applied in machine-learning and statistics to avoid overfitting and overly optimistic error estimates.
We employ a decision tree builder using the variance of explanatory variables and tree pruning using reduced-error pruning with back fitting (REPTree) implemented in the Weka package \cite{Weka}.

\section*{Acknowledgements}

This work was partly supported by PRESTO and by the ``Materials Research by Information Integration'' Initiative (MI$^2$I) project of the Support Program for Starting Up Innovation Hub, both from the Japan Science and Technology Agency (JST), Japan; by the Elements Strategy Initiative Project under the auspices of MEXT; and also by 
MEXT as a social and scientific priority issue (Creation of New Functional Devices and High-Performance Materials to 
Support Next-Generation Industries; CDMSI) to be tackled by using a post-K computer. We thank the Numerical Materials Simulator at NIMS to execute the systematic simulations in this study.

\section*{Author contributions statement}
T.L Pham and H. C. Dam developed the OFM descriptors and performed decision tree regression and KRR regression analyses. H. Kino performed DFT calculations. K. Terakura, T. Miyake, and H. Kino analyzed experiments on prediction of local magnetic moments. I. Takigawa and K. Tsuda performed analyses with coulomb matrix descriptors.  All authors reviewed the manuscript. 

\end{document}